\documentclass[12pt]{article}
\pdfoutput=1
\usepackage{authblk}
\usepackage{bm}
\usepackage{mathrsfs}
\usepackage{amsbsy}
\usepackage{graphicx}
\usepackage{subfigure}
\usepackage{amsmath}
\usepackage{amssymb}
\usepackage{braket}
\usepackage[numbers,sort&compress]{natbib}
\usepackage{lipsum}
\usepackage[]{hyperref}
\newcommand\blfootnote[1]{
\begingroup
\renewcommand\thefootnote{}\footnote{#1}
\addtocounter{footnote}{-1}
\endgroup
}

\newcommand{\mpref}[1]{Fig.\ref{#1}}
\numberwithin{equation}{section}

\begin{document}
	
	\begin{center}
		{\bf Entanglement Entropy, Phase Transition, and Island Rule for Reissner-Nordström-AdS Black Holes}\\

		\vspace{1.6cm}
		{\textbf{Shu-Yi Lin}$^{1}$,~ \textbf{Ming-Hui Yu}$^{1}$, ~\textbf{Xian-Hui Ge}$^{1,2,*}$}\blfootnote{* Corresponding author. gexh@shu.edu.cn}, \textbf{Li-Jun Tian}$^{1}$
		\vspace{0.8cm}
		
		$^1${\it Department of Physics, Shanghai University, Shanghai 200444, P.R. China} \\
            $^2${\it Institute for Quantum Science and Techenology, Shanghai University, Shanghai, 200444, P.R. China}

		\vspace{1.6cm}

		\begin{abstract}
This study focuses on the examination of the island rule within the context of four-dimensional Reissner-Nordström-AdS (4D RN-AdS) black holes, illuminating the intricate relationship between the entanglement entropy and phase transitions of black holes. The entanglement entropy of 4D RN-AdS black holes follows the anticipated linear growth pattern before ultimately declining to a constant value, in accordance with the well-established Page curve. The novelty of this study lies in the examination of the influence, previously unexplored, of the first-order phase transition on the shape and evolution of the Page curve in situations involving both eternal and evaporating black holes. Despite the morphological alterations of the curve induced by the transition, the inherent unitarity of the system persists. As the evaporation progresses, the Page curve displays diverse configurations, unveiling phenomena that are novel and defies traditional expectations, thereby enriching our comprehension of the thermodynamics of black holes interlinked with quantum information.
		\end{abstract}
	\end{center}
\newpage
\tableofcontents
\newpage

\section{Introduction} \label{introduction}
\qquad The black hole information paradox \cite{hawking1976breakdown} is one of the most fundamental problems to be solved in theoretical physics. Its origin lies in our comprehensive understanding of quantum gravity, and the resolution of the information paradox is fundamentally dependent on a thorough comprehension of quantum gravity. To elucidate this concept, we consider a scenario in which the collapse of a star into a black hole leads to an initial radiation state represented by the quantum state $\left | \Psi \right \rangle$. By means of the gravitational path integral, this initial state $\left | i \right \rangle$ transitions into the final state $\left | j \right \rangle$ , a process that encompasses Hawking radiation \cite{hawking1975particle,hawking1976breakdown,almheiri2021entropy}. To compute the von Neumann entropy of the black hole radiation, one can utilize the replica trick, which involves considering $n$ copies of the system, to compute the Renyi entropy, then taking the $n \to 1$ limit. The Renyi entropy is expressed as
\begin{equation}
    S=-\lim_{n \to 1} \frac{1}{n-1}\log{\text{Tr}\left(\rho^n\right)},
\end{equation}
where $\rho$ as the density matrix of either the black hole or the radiation, and it is possible to compute the Renyi entropies in a way that has a good continuation in $n$. The density matrix can be represented as $ \rho = \left | \Psi \right \rangle \left \langle \Psi \right | $, and the matrix elements $ \rho_{ij} $  as  $ \left \langle i | \Psi \right \rangle \left \langle \Psi | j \right \rangle $. For $ n \neq 1 $, the interior of the black hole can exhibit diverse connections among the $n$ replicas \cite{giddings2020wormhole}. At the early times, the Hawking saddle dominates the process, it would lead to the Hawking curve, and the density matrix $\rho$ is a thermal state or a mixed state. While at the late times, the replica wormholes saddle dominates the process, leading to the unitary Page curve \cite{almheiri2020replica,penington2022replica}. The state of the black hole corresponds to a pure state of density matrix $\rho$.
\par The complete entanglement entropy curve of a unitary evaporating black hole follows the Page curve \cite{page1993information}, as recently solved by the island rule \cite{penington2020entanglement,almheiri2019entropy,almheiri2020page,almheiri2019islands}. The Page curve describes the entropy of Hawking radiation, arising from the interplay between two saddle solutions: the Hawking saddle and the replica wormholes saddle \cite{marolf2020transcending}. The former is characterized by the von Neumann entropy, $S_{bulk}$, which shows a perpetual linear increase, reflecting the ongoing complexity within Hawking's computation \cite{hawking1974black}. The latter reflects the entropy defined by the island rule, associated with Renyi entropy or the Bekenstein-Hawking entropy. According to Page’s theorem, the entropy of a bipartite quantum system in a random pure state is given by \cite{page1993average,page1993information}
\begin{equation}
    S(R)=\min(\log d_R, \log d_B),
\end{equation}
where $d_R$ and $d_B$ denote the dimensions of the Hilbert space for the radiation and the black hole, respectively \cite{marolf2020transcending,ge2005reconsidering}. This relation holds when either $d_R \gg  d_B$ or $d_B \gg d_R$. Specifically, $\log d_R$ represents the entropy of the radiation, appropriately regularized up to a certain time, while $\log d_B$ corresponds to the Bekenstein-Hawking entropy of the black hole. Initially, the entropy increases due to the dominance of Hawking saddle, marked by the emission of additional radiation. After the Page time, a shift to the replica wormholes saddle occurs, aligning entropy with the Bekenstein-Hawking value and subsequently declining. This transition guarantees the unitarity of the black hole evaporation process, in accordance with the progression of the Page curve. Notably, similar behavior is observed in the $(3+1)$-dimensional Reissner-Nordstr{\"o}m-AdS black hole, capture the dominant s-wave sector, connected to non-gravitational baths within JT gravity \cite{penington2022replica,almheiri2020replica}. This approach applies to almost black holes \cite{penington2020entanglement}.
\par The Lorentzian interpretation of the replica wormholes saddle emphasizes that the entropy $S(R)$ encompasses not only the entropy of quantum field theory (QFT) within the primary region $R$, but also includes contributions from additional regions behind the black hole horizon, known as entanglement islands \cite{colin2021real,marolf2021observations}. These islands exist in the bulk of spacetime, where the metric fluctuates. In the semi-classical limit, the boundary of island, denoted as $\partial I$, corresponds to the Quantum Extremal Surfaces (QESs), which dynamically extremize the generalized entropy \cite{ryu2006holographic,hubeny2007covariant,faulkner2013quantum,engelhardt2015quantum}
\begin{equation}
    S_{I}(R)=\underset{\partial I}{\text{ext}} \left \{ \sum_{\partial I}\frac{\text{Area}(\partial I)}{4G_N} +S_{\text{QFT}}(R \cup I)  \right \}.
\end{equation}
The QESs can be points in $(1+1)$D or $2$D spatial surfaces in $(3+1)$D spacetime. The entropy of the primary region $R$ is determined by minimizing over the contributions from the island saddle 
\begin{equation}
    S(R)=\underset{I}{\text{min}}S_{I}(R) .
\end{equation}
This intricate interplay between entanglement islands and the replica wormholes saddle elucidates the behavior of black hole entropy and its connection to the underlying quantum structure. When the replica wormhole dominates, the island will emerge at the late times, thus the black hole information paradox will be solved hopefully.
\par Recent research on the black hole information paradox and the Page curve \cite{ge2003entropy,ge2005reconsidering,hashimoto2020islands,wang2021islands,guo2023page,yu2023page,yu2022island,yu2021page,yu2023entanglement,Yu:2024fks,Ahn:2021chg,Jeong:2023lkc,Geng:2024xpj} suggests that isolated regions, or “islands,” emerge during the evaporation of typical black hole models. Additionally, some work has been conducted on phase transitions \cite{cheng2014anisotropic,ghosh2019contact}. Observations indicate that phase transitions within the eternal black hole of JT gravity influence the deformation of the Page curve \cite{lu2023page,jackiw1985lower,teitelboim1983gravitation,cao2021thermodynamic}. This leads to a deviation from the behavior initially proposed by Page \cite{page1993information}, characterized by a non-monotonic trend rather than a straightforward increase and decrease. However, the investigation of the evaporating scenario remains incomplete.
\par In this paper, we exclude the extremal black hole situation and focus on the entanglement entropy of the non-extremal eternal 4D RN-AdS black hole, following the island rule. Our findings indicate that the island rule  resolves the paradox effectively. This outcome is plausible as the entanglement entropy demonstrates that the evaporation process of black hole is unitary and respect with the principles of quantum theory. Subsequently, we investigate the phase transitions specific to the 4D RN-AdS black hole and their influence on the Page curve. In the case of neutrality, the RN-AdS black hole transitions to a 4D Schwarzschild-AdS black hole, where the Hawking-Page phase transition occurs, and the Page curve maintains its standard form. Furthermore, we investigate the Page curve in the context of an eternal black hole to determine how varying charges, $Q$, influence its structure. Our analysis reveals that as the charge approaches the critical value, $Q_c$, the effects of the phase transition become less pronounced. We also model the evaporating RN-AdS black hole under the assumption of coupled baths or the removal of the baths to ensure evaporation rather than an eternal black hole state. This approach allows us to observe the Page curve behavior during black hole evaporation. In the charged case, the Page curve displays a variety of behaviors—monotonic, non-monotonic, or even a discontinuity—depending on the event horizon radius. Importantly, these variations do not compromise the unitarity of black hole evaporation.
\par The structure of this paper is organized as follows: in section \ref{Review of RN-AdS black hole}, we provide a brief review of the RN-AdS black hole. In Section \ref{The entanglement entropy in 4-dimensional RN-AdS black hole}, we employ the island formula to compute the entanglement entropy for situation without island and with an island, and then we dedicate to illustrating the Page curve and determining the scrambling time. The effects of phase transitions on the Page curve, influenced by varying the value of charges Q, are explored in Section \ref{Phase transitions and page curve}. Finally, we presents our conclusions in section \ref{Discussion}.

\section{Review of RN-AdS black hole} \label{Review of RN-AdS black hole}
\qquad We first review some properties of RN-AdS black hole. The RN-AdS black holes in 4D spacetime are characterized by the action in the form of \cite{chamblin1999charged}:
\begin{equation}
    I=\frac{1}{16 \pi G_{N}} \int d^{4} x \sqrt{-g}\left(R+\frac{6}{\ell ^{2}}- F_{\mu \nu} F^{\mu \nu}\right),
\end{equation}
where $\ell$ is the AdS length scale, defined by $L \equiv \sqrt{-3 \Lambda }$ with $\Lambda$ being the cosmological constant, $R$ is the Ricci scalar, and $g$ is the dimensionless coupling constant of U(1) gauge field. The metric of 4D static charged RN-AdS black hole is
\begin{equation}
    d s^{2}=-f(r) d t^{2}+\frac{d r^{2}}{f(r)}+r^{2}(d \theta^2 +\sin^2 {\theta} d\phi ^2). 
\end{equation}
Setting the Newton's constant and the Coulomb constant equal to 1, i.e., $G_N=K=1$, the function $f(r)$ is defined as follows:
\begin{equation}
    f(r)=1-\frac{2M}{r}+\frac{Q^2}{r^2}+\frac{r^2}{\ell ^2},
    \label{eq:f(r)}
\end{equation}
where $M$ and $Q$ in the metric function are the mass and charge of the black hole, respectively. 
\par One can obtain the Hawking temperature, Bekenstein-Hawking entropy, and the electric potential by deriving the first law, which can be written as follows:
\begin{equation}
    \begin{aligned}
        T & =\frac{1}{4\pi r_+}\left ( 1+\frac{3r_{+}^{2}}{\ell ^2}-\frac{Q^2}{r_{+}^{2}} \right ) ,\\
        S & =\pi r_{+}^{2},\\
        \phi & = \frac{Q}{r_+}.
    \end{aligned}
\end{equation}
The event horizon is located at $r=r_+$. The mass of the black hole can be derived with the condition $f(r_+)=0$, i.e.:
\begin{equation}
    M=\frac{r_+}{2}+\frac{\phi Q}{2}+\frac{r_+^3}{2 \ell ^2}.
\end{equation}
Clearly, the mass of the black hole is a function of the event horizon radius $r_+$, charge $Q$, and electric potential $\phi$.
\par In AdS spacetime, an infinitely large effective potential exists at the boundary of spacelike infinity. This potential reflects the Hawking radiation emitted by AdS black holes back towards them, effectively transforming the AdS spacetime into an infinite potential well. In the case of AdS black holes with sufficiently large mass, the reflected Hawking radiation can return to the black hole rapidly enough to establish a thermal equilibrium. This results in a stable black hole with a non-zero temperature that does not undergo evaporation. To induce evaporation in the black hole, we introduce a coupling with a non-gravitational bath at finite temperature, as described by Almheiri et al. in \cite{almheiri2019entropy}. Our study focuses on an AdS black hole that has reached equilibrium with a bath located at the boundary, making the black hole evaporate. One can refer to \mpref{fig:1}.
\par Then the black hole could evaporate in this way. In order to obtain the maximally extended double-sided geometry and to better facilitate subsequent calculations, we perform the Kruskal coordinate transformation. The definition of the tortoise coordinate is
\begin{equation}
    r_{\ast}(r)=\int^r \frac{1}{f(r)}dr,
\end{equation}
and define the Kruskal coordinate are defined as:
\begin{equation}
    U=-e^{-\kappa (t-r_*)}, \qquad V=+e^{\kappa (t+r_*)},
\end{equation}
where $\kappa$ is the surface gravity, which can be written as:
\begin{equation}
    \kappa =\frac{1}{r_+^2}\left(M-\frac{Q}{r_+}\right)+\frac{r_+}{l^2}.
\end{equation}
We refer to the Kruskal-Szekeres coordinate in the RN-AdS spacetime which given by
\begin{equation}
    \begin{aligned}
        ds^2 & =-\Omega ^2 (r)  dUdV \\
             & =-\Omega ^2 (r)\kappa ^2 e^{2\kappa  r_*}dt^2+\Omega ^2 (r)  \kappa ^2 e^{2\kappa r_*}f^{-2}(r) dr^2,
        \label{eq:conformal}
    \end{aligned}
\end{equation}
comparing equation \eqref{eq:f(r)} and equation \eqref{eq:conformal}, we obtain the conformal factor $\Omega (r)$ in bulk spacetime \footnote{After here, we set the AdS length $\ell =1$ for convenience.}  
\begin{equation}
    \begin{split}
        \Omega^2 (r) & =\frac{1}{\kappa ^2 e^{2 \kappa r_*}}f(r) \\
                     & =\frac{\left(-Q+Mr_++r^4 \right)\left(Q^2+r(-2M+r+r^3) \right)}{r^5}.
    \end{split}
\end{equation}
Due to the fact that bath is in a Minkowski spacetime, characterized by the metric $ds^2 =-dt^2+dr^2$, in the flat thermal bath region, this conformal factor is
\begin{equation}
    \Omega ^2 (r) =\frac{1}{\kappa ^2 e^{\kappa r}}.
    \label{eq:conformal factor}
\end{equation}
In this way, the black hole is able to radiate, allowing the asymptotic observer to collect Hawking radiation in the bath region. Consequently, the island formula can be employed to calculate the entanglement entropy of the Hawking radiation.

\section{Entanglement entropy in 4-dimensional RN-AdS black hole} \label{The entanglement entropy in 4-dimensional RN-AdS black hole}
\qquad In this section, we calculate the entanglement entropy of 4D RN-AdS black hole in two different situation: one without island, and the other with island during the black hole’s late-stage evaporation. The island formula is given by
\begin{equation}
    S(R)=\text{min}\left \{ \text{ext}\left [ \frac{\text{Area}(\partial I)}{4G_N} +S_{\text{Bulk}}(R \cup I)   \right ]  \right \} .
    \label{Island}
\end{equation}
For simplicity, we only consider the single island case, as multiple islands contribute only to sub-leading terms and do not affect our main results. Consequently, our analysis will concentrate on the emergence of a single island at the final stage of evaporation.

\par We assume that the entire system is in a pure state at $t=0$. We set the coordinates of these two points to be $(t,r)=(t_b,b)$ and $(-t_b-i \beta /2,r)=(t_b,b)$, respectively. According to the complementarity principle of von Neumann entropy, the entropy of the conformal field theory (CFT) in the region outside $(-\infty ,b_-)\cup (b_+,+\infty )$ is equivalent to the entropy of the radiation region. Correspondingly, the entanglement entropy of the radiation is given by:

\subsection{Without Island}  \label{Without island}
\begin{figure}[htb]
\centering
\includegraphics[scale=0.24]{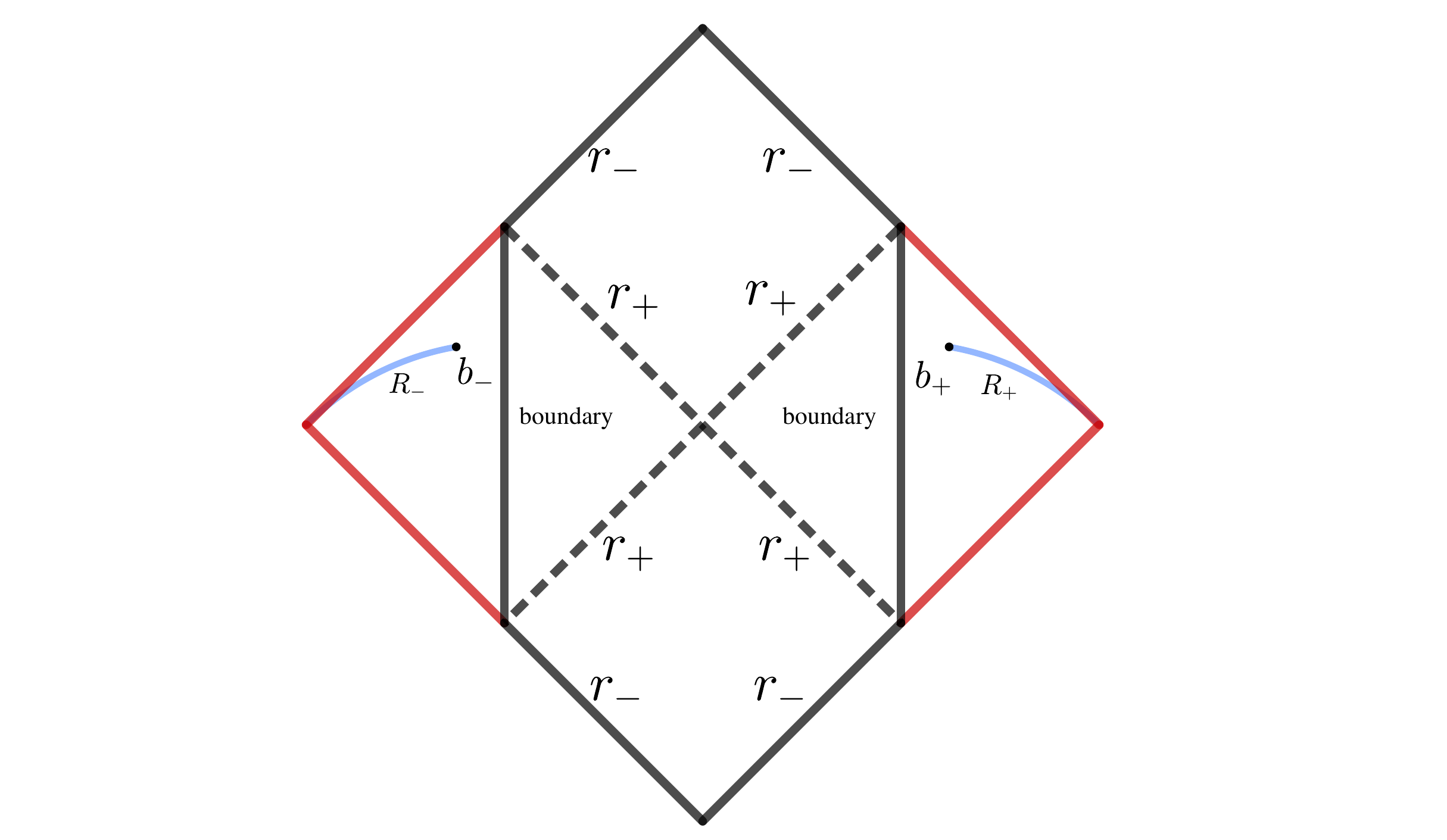}
\caption{The Penrose diagram of an eternal 4D $\text{RN-AdS}_{4}$ black hole without island. The black region represents the black hole, and the red region denotes the bath region. The boundary conditions are imposed on the boundary. $R_{\pm}$ indicate the radiation regions on the right and left wedges, and $b_{\pm}$ are the boundaries of the radiation regions $R_{\pm}$. } 
\label{fig:1}
\end{figure}

\qquad First, we calculate the entanglement entropy without island at early times. Due to the divergent nature of entropy in higher dimensions, and the absence of a definitive formula for its calculation, we adopt a large distance limit to validate the s-wave approximation which neglect the spherical part of the metric and only considers the conformally flat part. Thus the Hawking radiation can be described properly by a long distance observer. In this paper, we exclusively focus on the neutral uncharged radiation from the black hole and disregard the Schwinger effect. This configuration is shown in \mpref{fig:1}. 
\par We assume that the entire system is in a pure state at $t=0$. We set the coordinates of these two points are $(t,r)=(t_b,b)$ and $(-t_b-i \beta /2,r)=(t_b,b)$, respectively. According to the complementarity principle of von Neumann entropy, the entropy of the conformal field theory (CFT) in the region outside $(-\infty ,b_-)\cup (b_+,+\infty )$ is equivalent to the entropy of the radiation region. Correspondingly, the entanglement entropy of the radiation is given by
\begin{equation}
    S_{\text{Bulk}}=S(R)=\frac{c}{3}\log[d(b_+,b_-)].
\end{equation}
The $d(b_+,b_-)$ is the geodesic distance between the point $b_{\pm}$. Then, we have
\begin{equation}
    \begin{aligned}
        d(b_+,b_-)^2&=\Omega(b_+)\Omega(b_-)\left [ U(b_-)-U(b_+) \right ]\left [ V(b_+)-V(b_-) \right ]   \\
                    & = \Omega^2(b)\left [ U(b_-)-U(b_+) \right ]\left [ V(b_+)-V(b_-) \right ]  ,
    \end{aligned}
\end{equation}
where the conformal factor $\Omega ^2 (b)$ could be written as \eqref{eq:conformal factor}
\begin{equation}
    \Omega ^2(b)=\frac{1}{\kappa ^2 e^{2\kappa b}}.
\end{equation}
Therefore, the entanglement entropy without island is given by
\begin{equation}
\begin{aligned}
     S_{Rad} &= \frac{c}{6} \log{\left [ \frac{1}{\kappa ^2 e^{2\kappa b}} 2e^{2\kappa b}\left [ \cosh(2\kappa t_b)+1 \right ]  \right ] }\\
     &= \frac{c}{6}\log \left [ \frac{4}{\kappa ^2 }\left [ \cosh ^2(\kappa t_b) \right ]  \right ] .
\end{aligned}
\end{equation}
For late times ($t_b\to \infty$), the above equation simplifies to 
\begin{equation}
    \begin{aligned}
        S_{Rad} &\simeq \frac{c}{6}\log\left(\frac{4}{\kappa ^2}\frac{1}{4}e^{2\kappa t_b}\right)=\frac{c}{6}\log \left(\frac{1}{\kappa^2}e^{2\kappa t_b}\right)\\
        &\simeq \frac{c}{3}\kappa t_b ,
        \label{eq:without island}
    \end{aligned}
\end{equation}
which exhibits linear growth in time. Therefore, at late times, the entropy of radiation will exceed the Bekenstein-Hawking entropy of the black hole, leading to eventual loss of information from the black hole due to the absence of an $S$ matrix that ensures unitary evaporation. Such a situation presents a contradiction with the finite von Neumann entropy of a finite-dimensional black hole system, implying that the black hole would evolve to a mixed state, thereby violating the principle of unitarity. It is shapen the black hole information paradox. To solve this issue, we propose the inclusion of an island in our subsequent calculations. We expect that the emergence of an island could resolve these inconsistencies and restore unitarity in black hole evaporation.

\subsection{With Island}  \label{With island}
\begin{figure}[htb]
\centering
\includegraphics[scale=0.23]{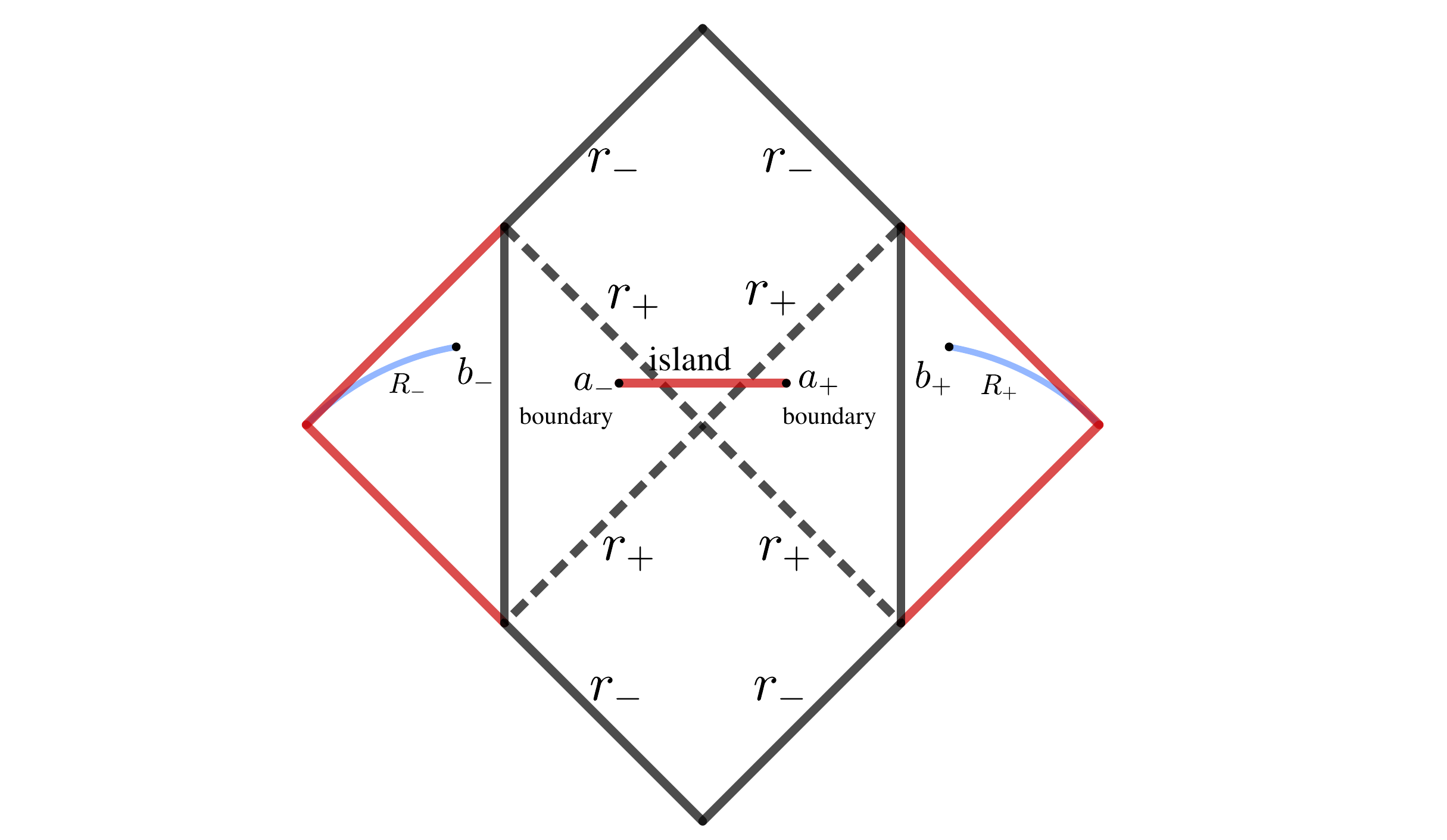}
\caption{The Penrose diagram of an eternal 4D $\text{RN-AdS}$ black hole with an island. $R_{\pm}$ are the radiation regions on the right and left wedges, and $b_{\pm}$ are the boundaries of the radiation region $R_{\pm}$. The boundary of the island is denoted as $a_{\pm}$. }
\label{fig:2}
\end{figure}

\qquad In the subsequent analysis, we consider another modified gravitational method, i.e., the proposal of quantum entanglement islands. Although entanglement entropy generally depends on the cut-off surface\footnote{The cut-off surface is artificially selected, and it is generally stipulated that Hawking radiation is emitted from the cut-off surface by considering the back-reaction of radiation. For a 4D Schwarzschild black hole, the cut-off surface is approximately $2-3$ times of the Schwarzschild radius, a region where gravitational effects can be considered negligible.} scale, in 2D systems with conformal symmetry, one can employ regularization and renormalization techniques to obtain a well-defined entanglement entropy. This approach was pioneered by Holzhey, Larsen, and Wilczek \cite{holzhey1994geometric}, and later refined by Bianchi and Smerlak \cite{bianchi2014entanglement}. From this, we can derive the formula for fine-grained entropy. We calculate the entanglement entropy with an island at the late stage of black hole evaporation. The entanglement entropy for this disconnected interval $R \cup I$ (see \mpref{fig:2}) as follows \cite{hashimoto2020islands}: 
\begin{equation}
    S_{\text{bulk}} (R \cup I)=\frac{c}{3} \log \frac{d(a_+,a_-)d(b_+,b_-)d(a_+,b_+)d(a_-,b_-)}{d(a_+,b_-)d(a_-,b_+)}.
\end{equation}
\par We define the coordinates of the island as $a_{\pm} = (\pm t_a, a)$. Using the Kruskal coordinates and applying the conformal factor $\Omega (r)$, the generalized entanglement entropy in this time is
\begin{equation}
    \begin{aligned}
        S_{\text{gen}} & =\frac{2\pi a^2}{G_N}+ \frac{c}{6}\log\left [ \frac{16f(a)}{\kappa ^4 } \cosh ^2(\kappa t_a) \cosh ^2(\kappa t_b)  \right ]\\
          & +\frac{c}{3}\log \left [ \frac{\cosh \left [ \kappa (r_*(a)-r_*(b)) \right ]-\cosh\left [ \kappa (t_a-t_b) \right ]  }{\cosh \left [ \kappa (r_*(a)-r_*(b)) \right ]+\cosh \left [ \kappa (t_a+t_b) \right ]  }  \right ]  .
    \end{aligned}
    \label{eq:S general}
\end{equation}
\par Firstly, we consider the behavior of entanglement entropy at early times. We assume that the following approximation is valid
\begin{equation}
    t_a,t_b \simeq 0, \qquad  t_a, t_b\ll r_+, \qquad  t_a,t_b\ll 1/\kappa \ll r_*(b)-r_*(a).
\end{equation} 
We pick the cut-off surface far away from the horizon $b\gg r_+$. Then the last term in the generalized entropy can be properly omitted in this approximation. The generalized entropy is given approximately as follows
\begin{equation}
    \begin{aligned}
        S_{gen}(\text{early}) & \simeq \frac{2\pi a^2}{G_N}+\frac{c}{6}\log  \left [ \frac{16f(a)}{\kappa ^4}\cosh ^2(\kappa t_a)\cosh ^2 (\kappa t_b)  \right ] \\
                & \simeq \frac{2\pi a^2}{G_N}+\frac{c}{6} \left [ \log f(a)+\cosh{(2\kappa t_a) }\right ].
    \end{aligned}
    \label{eq:S}
\end{equation} 
\par In order to obtain the location of the island, we need to solve the following equation. We can solve the following system of equations by extremizing the generalizd entropy
\begin{equation}
    \begin{aligned}
        \frac{\partial S_{\text{gen}}(\text{early})}{\partial t_a}=\frac{c\kappa }{3}\tanh (\kappa t_a) = 0,\\
        \frac{\partial S_{\text{gen}}(\text{early})}{\partial a}=\frac{4 \pi a}{G_N}+\frac{c}{6}\frac{f'(a)}{f(a)} = 0.
    \end{aligned}
    \label{eq:partial St}
\end{equation}
The solution to the first equation is $t =0 $. For the second equation, employing approximation methods and assuming the higher-order terms of $\frac{a-r_+}{r_+}$ to be negligible, we deduce an effective solution $a \simeq r_+$ plus higher-order terms, which are smaller than the Planck length $l_p$. This result suggests that there is no quantum extremal surface that leads to the increase of entropy to the extremal point before the Page time. The entanglement entropy of the radiation only relies on the radiation itself in the early time and is not affected by the island. Subsequently, we extend our analysis at late times situation where an island emerges.
\par Now we turn to the construction with a single island. In this time, the construction of the matter entropy also grows, exhibiting a linear increase in fine-grained entropy during the late stages of black hole evaporation. This phenomenon is expected to occur around the Page time. There are some approximate equations as follows
\begin{equation}
    \begin{aligned}
        t_a,t_b\gg b> r_+, \quad  \cosh ( \kappa t )=(e^{\kappa t}+e^{- \kappa t})/2,\\
        \cosh {\kappa_+(t_a+t_b)}\gg \cosh{\kappa_+(r_*(a)-r_*(b))},\\
        \cosh{\kappa_+(r_*(a)-r_*(b))}\simeq \frac{1}{2}e^{\kappa_+(r_*(b)-r_*(a))}.
    \end{aligned}
\end{equation}
Thus the distance among the points, $a_{\pm}$ and $b_{\pm}$ will have following behavior as \cite{hashimoto2020islands}
\begin{equation}
    d(a_+,a_-) \simeq d(b_+,b_-) \simeq d(a_+,b_-) \simeq d(a_-,b_+)  \ge d(a_+,b_+) \simeq d(a_-,b_-).
    \label{eq:distance}
\end{equation}
Plugging equation \eqref{eq:distance} into equation \eqref{eq:S general}, the generalized entropy is given by
\begin{equation}
\begin{aligned}
        S_{\text{gen}}(\text{late}) & =\frac{2\pi a^2}{G_N}+ \frac{c}{6}\log\left [ \frac{16f(a)}{\kappa ^4}\cosh ^2(\kappa t_a) \cosh ^2(\kappa t_b)  \right ]\\
            &+\frac{c}{3}\log \left [ \frac{\cosh \left [ \kappa (r_*(a)-r_*(b)) \right ]-\cosh\left [ \kappa (t_a-t_b) \right ]  }{\cosh \left [ \kappa (r_*(a)-r_*(b)) \right ]+\cosh \left [ \kappa (t_a+t_b) \right ]  }  \right ] \\
            &\simeq \frac{2\pi a^2}{G_N}+\frac{c}{3}\log \left [\frac{4\sqrt{f(a)}}{\kappa^2} \left [ \cosh \left [ \kappa(r_*(b)-r_*(a))  \right ] -\cosh{\left [ \kappa(t_a-t_b) \right ]  }  \right ]  \right ]  \\
            & \simeq \frac{2\pi a^2}{G_N}+\frac{c}{3}\log \left [\frac{4\sqrt{f(a)}}{\kappa^ 2} \right ]-\frac{c}{3} ( r_*(a)-b ) -\frac{2c}{3}e^{-\kappa(b-r_*(a))} ,
    \end{aligned}
\end{equation}
where in the second approximate equation, we take the partial derivative of $S$ with respect to $t $
\begin{equation}
    \frac{\partial S_{gen}(\text{late})}{\partial t_a}= \frac{c}{3} \frac{\kappa \sinh{ \kappa (t_a-t_b)}}{\cosh \left [ \kappa (r_*(a)-r_*(b)) \right ]-\cosh \left [ \kappa (t_a-t_b) \right ]}=0.
    \label{eq:partial S}
\end{equation}
We find that the partial derivative equation \eqref{eq:partial S} is equal to zero only when $t_a \approx t_b $, which implies that the generalized entropy is independent of time. By solving the extremal condition $\partial S_{\text{gen}} / \partial a =0 $, we obtain
\begin{equation}
        \frac{4 \pi a}{G_N}=\frac{c}{3}  \frac{1}{f(a)} \left [ 2\kappa e^{-\kappa (r_*(b)-r_*(a))}+1 \right ] -\frac{c}{6}\frac{f'(a)}{f(a)}.
        \label{eq:located a}
\end{equation}
At late times and in the large distances limit \eqref{eq:distance}, we can take the near horizon limit, $a \simeq r_+$. We can approximate the tortoise coordinate $r_*$ by
\begin{equation}
    \begin{aligned}
        f(r) & \simeq f'(r)(r-r_+)=2\kappa (r-r_+),\\
        r^*(r) &=\int \frac{1}{f(r)}dr \simeq \frac{1}{2\kappa} \int \frac{dr}{r-r_+}\\
        &=\frac{1}{2\kappa}\log {\left | \frac{r-r_+}{r_+} \right | }.
    \end{aligned}
    \label{eq:r approximation}
\end{equation}
Substituting the approximation into equation \eqref{eq:located a}, we obtain the following result
\begin{equation}
    a=\frac{c}{4\pi G_N}\left(\frac{1}{2}-\frac{1}{2}\sqrt{1-\frac{8\pi G_N r_+}{c^2}\left(1+\frac{c}{6r}\left(e^{-b\kappa}+1\right)\right)}\right)
    \label{eq:ainitial}
\end{equation}
Performing a Taylor expansion, we obtain the final expression
\begin{equation}
    a \simeq r_{+}+\frac{c^2G_N^2}{12^2 \pi r_+^{3}} e^{-2\kappa b}+\mathcal{O} (G_N ^3).
    \label{eq:a value}
\end{equation}
By analyzing the expression, we can derive the entanglement entropy
\begin{equation}
    \begin{aligned}
        S &=\frac{2 \pi r_+ ^2}{G_N}+\mathcal{O} (G_N ^3)\\
        & \simeq 2S_{BH}.
    \end{aligned}
    \label{eq:with island}
\end{equation}
The second term has a finite value. Therefore, at late times, the dominant contribution to entanglement entropy is provided by the Bekenstein-Hawking entropy. The entanglement entropy of an eternal black hole will remain constant during its evaporation, ensuring the unitarity of the final state. This approach is more reasonable and addresses our problem in the result \eqref{eq:without island} in the final step. Consequently, the island solution is consistent with the Page curve prediction, indicating that Hawking radiation is unitary and information is preserved in the 4D RN-AdS black hole.

\subsection{Page curve and scrambling time} \label{Page curve and scrambling time}
\qquad In this section, we calculate the Page time and scrambling time. Based on the value of entanglement entropy at early times and late times, we can plot the Page curve.
\par
\begin{figure}[htbp]
\centering
\includegraphics[scale=0.5]{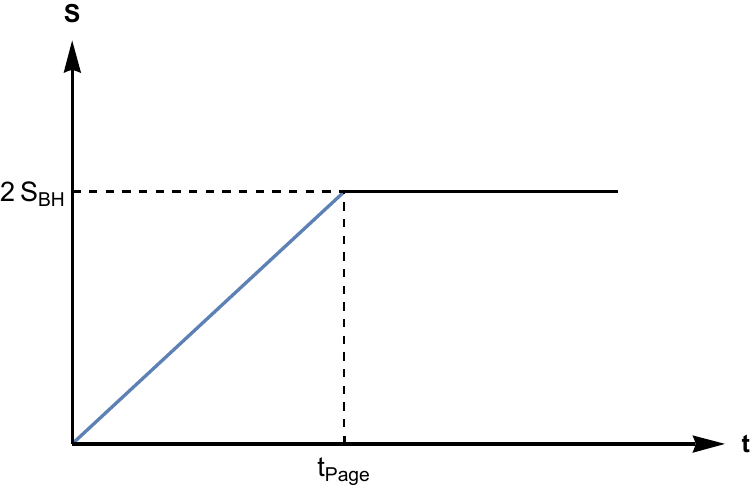}
\caption{Page curve of 4D RN-AdS black hole. The blue line represents entanglement entropy at the early times, while the black line represents entropy at late times. They intersect at the point in 2$S_{\text{BH}}$ at time $t_{\text{Page}}$.}
\label{fig: Page curve}
\end{figure}

 Page time is the time when entropy grows to its maximal value. At early times, the dominant term is as shown in equation \eqref{eq:without island}, while later, the term in equation \eqref{eq:with island} becomes larger than the time-dependent term. We estimate the Page time by identifying the point of intersection between the entropy curve without the island at very early times and the entropy curve with the island at late times. We determine the Page time by equating equation \eqref{eq:without island} and equation \eqref{eq:with island}. Letting them be nearly equal during the late times and using the Hawking temperature $T_H = \frac{\kappa}{2\pi}$, we can perform the calculation as follows
  \begin{equation}
    \begin{split}
        t_{\text{Page}}&  \simeq \frac{6S_{BH}}{c \kappa }\\ 
                &  \simeq \frac{3S_{BH}}{c \pi T_H}.
    \end{split}
  \end{equation}
According the location of island, we find that the island emerges near the horizon. Then, we obtain the Page curve as \mpref{fig: Page curve}.
\par The scrambling time is the shortest time for an outside observer to recover the information in the Hawking radiation that fell into the black hole before \cite{hayden2007black,wang2021islands}. It approximate through equations \eqref{eq:r approximation} and \eqref{eq:a value}, and is given by
  \begin{equation}
    \begin{aligned}
        t_{scr} & \equiv  \text{Min}  [ \Delta t ] =r_*(b)-r_*(a)   = b-\frac{1}{2\kappa}\log {\left | \frac{c^2G_N^2}{144 \times 4 \pi r_+^4}e^{-2\kappa b} \right | }\\
        & \simeq \frac{1}{\kappa}\log {\left ( \frac{12 \times 2 \pi r_+^2}{cG_Ne^{-\kappa b}} \right ) } \\
        & \simeq \frac{1}{ \kappa} \log{ S_{BH}} ,
    \end{aligned}
    \label{eq:scr t}
  \end{equation}
where we have used the equation \eqref{eq:a value} to reduce, and use the near horizon approximation, and $r_+$ is much larger than other terms, so the logarithm of $r_+$ is almost the same as $r_+$ in the third line.

\section{Phase transitions and Page curve} \label{Phase transitions and page curve}
\qquad In our previous discussions, we have analyzed the situations both without island and with an island in the context of entanglement entropy. Our next objective is to investigate the possibility of a first-order phase transition occurring in the RN-AdS black hole, a phenomenon that could potentially influence the entanglement entropy of evaporating black hole. This consideration is integral to our analysis, and henceforth, this section is dedicated to examining the Page curve as affected by the phase transitions of the black hole.
\par Initially, we ascertain the phase transition by evaluating the free energy of the RN-AdS black hole. The free energy $\cal{F}$ can be obtained by computing the Euclidean action in the semiclassical approximation, and it is given by
\begin{equation}
    \begin{aligned}
        {\cal F} (T,V)&=M-TS\\
        &=\frac{1}{2}\left(r_+-2\pi Tr_+^2+\frac{Q^2}{r_+} \right).
    \end{aligned}
\end{equation}
\par Subsequently, we plot the ${\cal F}-T$ diagram, referenced in \mpref{Q=01}, and compare it to the temperature-radius $T-r_+$ diagram, as detailed in \mpref{Q=02}. It becomes conspicuously evident that in the $Q=0$ case, only the Hawking-Page phase transition occurs. This phenomenon signifies that a black hole with larger free energy is more unstable. Consequently, in such a case, only a solitary black hole exists. Therefore, when the charge $Q=0$, the Page curve is unaffected by the Hawking-Page first-order phase transition, as shown in \mpref{fig:Sch}. 
\par Further, we present the $T-r_+$ diagram and the ${\cal F}-T$ diagram, as depicted in \mpref{fig:Q=0.03}, for the $Q \neq 0$ case. Here, we consider eternal black holes with varying charges $Q$. It becomes evident that as the charge $Q$ approaches the critical charge, which permits phase transitions, its impact on the Page curve nearly diminishes. Delving deeper into this effect, we focus on the most pronounced case where the charge $Q$ is small. 
\par Drawing inspiration from the aforementioned results, we simulate the evaporation of the RN-AdS black hole under scenarios involving coupled baths or the absence thereof. This allows us to determine the effects on the evaporating black hole, particularly how these different conditions influence the black hole's temperature, entropy, and the Page curve. By comparing these cases, we aim to deepen our understanding of black hole thermodynamics and information preservation during the evaporation process. Additionally, we discuss whether phase transitions occurring before or after the Page time induce any alterations in the Page curve. We uncover an intriguing phenomenon—the behavior of the Page curve, whether monotonic, non-monotonic, or discontinuous, is predicated on the initial event horizon radius $r_+$. The intricate details of this analysis are delineated in the following subsection.

\subsection{Neutral case}
\begin{figure}[htb]
\centering
\subfigure[\scriptsize{}]{\label{Q=01}
\includegraphics[scale=0.5]{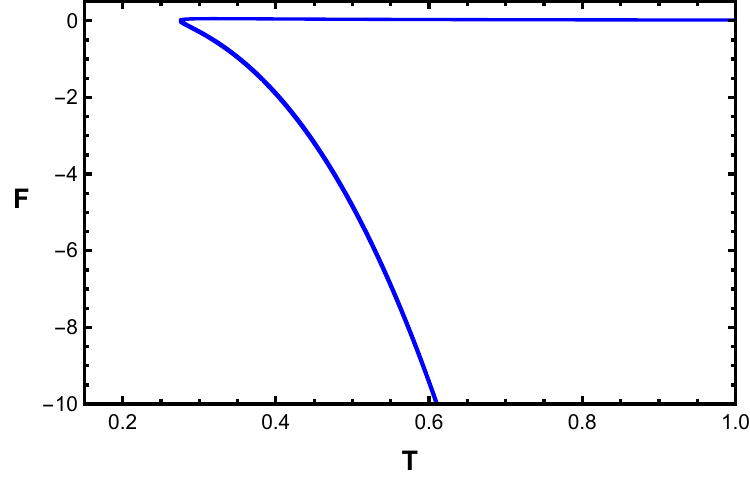}
}
\quad
\subfigure[\scriptsize{}]{\label{Q=02}
\includegraphics[scale=0.5]{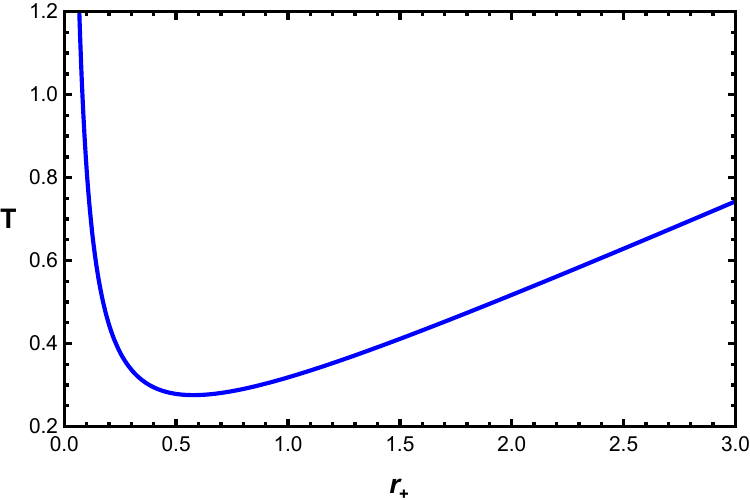}
}
\caption{The ${\cal F}-T$ figure for Schwarzschild black holes. (a) On the left, for $Q=0$, the $T-{\cal F}$ Figure. $\cal F$ is depicted as functions of $T$ when $Q=0$. As $r_+$ increases, the temperature corresponds to two distinct values of free energy within the range $0.276<T<0.609$. This implies the presence of only one black hole in this case, as the branch with the lower free energy is more stable. (b) On the right, for $Q=0$, the $T-r_{+}$ diagram. Each $r_+$ corresponds to a unique temperature $T$. $T$ is a function of $r_+$, and as $r_+$ becomes large, the temperature also increases significantly.}
\label{fig:Q=0}
\end{figure}

\qquad First, we observe the case where the charge $Q=0$, reducing the RN-AdS black hole to a Schwarzschild-AdS black hole. The 4D Schwarzschild black hole has only one horizon, labeled as $r_+$. The free energy is given by 
\begin{equation}
    {\cal F} = \frac{1}{2}\left(r_+-2\pi Tr_+^2 \right),
\end{equation}
and the temperature can be written as
\begin{equation}
    T=\frac{1}{4\pi r_+}(1+3r^2_+).
\end{equation}
According to \cite{hawking1983thermodynamics, eom2022hawking}, a Hawking-Page phase transition occurs between an AdS black hole with radiation and thermal AdS. To illustrate this, we can depict the free energy, $\cal F$, as a function of the temperature $T$, and similarly, the radius $r_+$ as a function of the temperature $T$. As illustrated in \mpref{fig:Q=0}, when the black hole temperature $T>0.276$, at the fixed temperature, there exit two sizes of black holes.
The ${\cal F}-T$ curve shows that there are two values of free energy at the same temperature. This indicates that there exists only one more stable black hole with a larger black hole radius, because when the temperature is fixed, the lower free energy, which is related to the black hole, is more stable.
\par
\begin{figure}[htbp]
\centering
\includegraphics[scale=0.5]{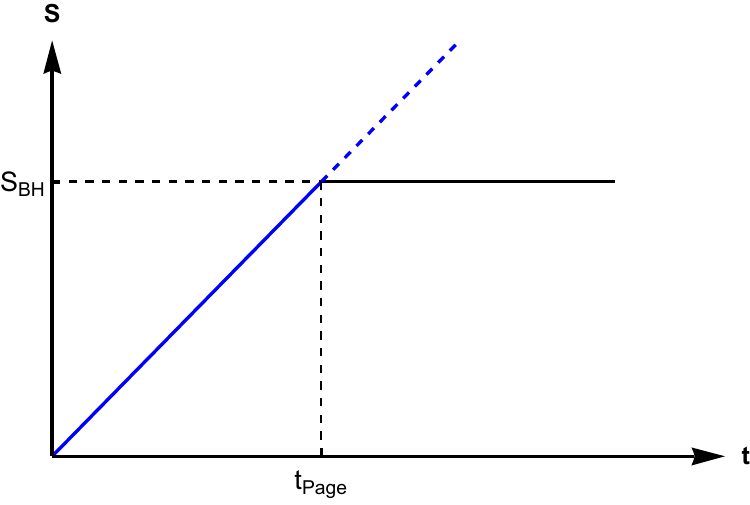}
\caption{Page curve of 4D Schwarzschild black hole. The Page curve is delineated by a blue line representing the entanglement entropy in the early stages, and a black line indicating the entropy at later stages. These lines intersect at a point designated as $S_{\text{2BH}}$, corresponding to the pivotal time $t_{\text{Page}}$. Notably, this convergence is unaffected by the Hawking-Page phase transition.}
\label{fig:Sch}
\end{figure}

 The entanglement entropy of the Schwarzschild black hole can be computed using equation \eqref{Island}, analogous to the method delineated in Section \ref{The entanglement entropy in 4-dimensional RN-AdS black hole}. Drawing parallels to the RN-AdS black hole, the equation governing the parameter retains the identical form as presented in equations \eqref{eq:ainitial} and \eqref{eq:a value}. The resultant entanglement entropy is articulated as
\begin{equation}
     S_{\text{Sch}} = \text{Min}{ \left [ \frac{2}{3} \pi T(r_{h}) t, \  \frac{4 \pi r_+^2}{G_N}  \right ] }.
\end{equation}
The Page curve, as depicted in \mpref{fig:Sch}, initially increases linearly and tends to level off after the Page time.

\subsection{Charged case}
\qquad In this charged case, we leverage the analogy between the RN-AdS black hole and the van der Waals fluid, as established in Ref \cite{kubizvnak2012p}, especially in the context of phase transitions. We examine how the Page curve is altered by varying the event horizon radius $r_+$.
\par
\begin{figure}[htb]
\centering
\subfigure[\scriptsize{}]{\label{Q=0.031}
\includegraphics[scale=0.5]{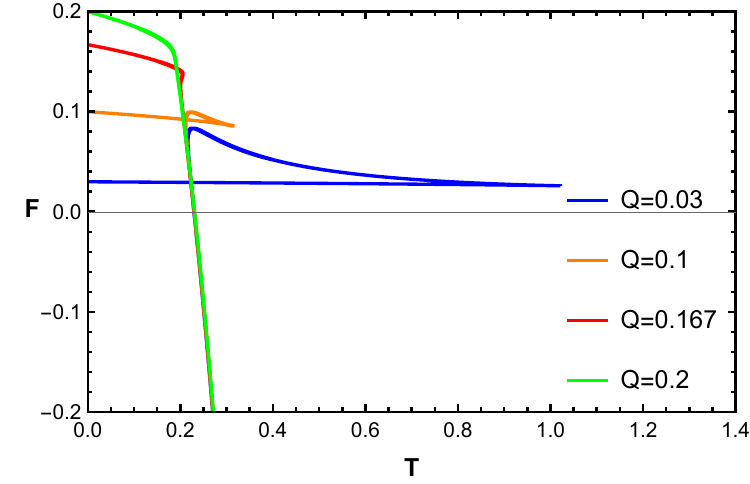}
\label{fig:F-T 1}
}
\quad
\subfigure[\scriptsize{}]{\label{Q=0.032}
\includegraphics[scale=0.5]{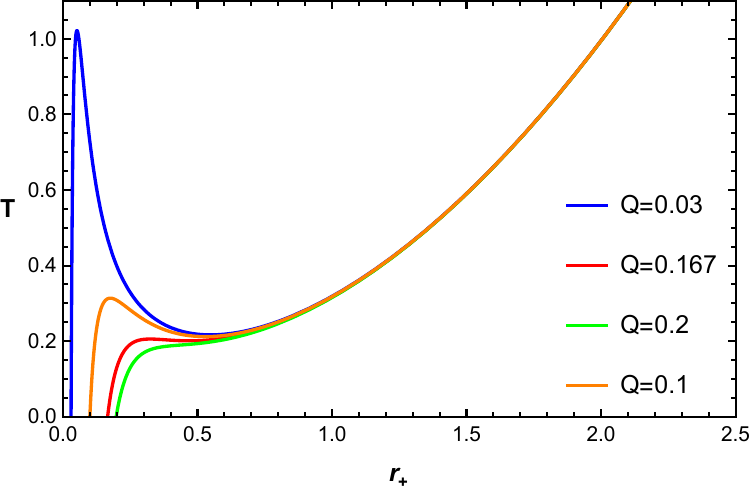}
\label{fig:r-T 1}
}
\caption{(a) The diagram for free energy $\cal F$ and temperature $T$ with non-vanishing charges. Here we set the charge smaller or larger than $Q_c$, as well as equal to the critical charge. When $Q<Q_c$, the ${\cal F}-T$ curve shows the characteristic swallowtail behavior, indicating an impending phase transition. Conversely, for $Q>Q_c$, the ${\cal F}-T$ curve remains smooth. (b) On the right, the $T-r_{+}$ figure with $Q=0.03$, 0.01, 0.167 and 0.2. At low temperatures, only one black hole exists. As the temperature increases and $Q<Q_c$, three distinct black holes emerge; however, due to the instability of the intermediate black hole, only a large and a small black hole remain; when $Q>Q_c$, there is no phase transition.}
\label{fig:Q=0.03}
\end{figure}

 Firstly, we plot the free energy $\cal F$ as a function of temperature $T$, as shown in \mpref{Q=0.031}. The ${\cal F}-T$ curves, colored blue and orange, display the characteristic swallowtail behavior when the charge $Q$ is below a certain value of charge, which is a candidate of the value of critical charge $Q_c$. Conversely, scenarios where $Q$ equals to or exceeds this certain value of charge are represented by smooth curves, depicted in red or green. This diagram indicates that a phase transition occurs for $0<Q<Q_c$, with the critical point being an inflection point. For $Q>Q_c$, it represents a thermodynamically stable branch with no phase transition structures. The blue and orange curves each have three branches: the lower branch corresponds to the large black hole (LBH), the upper branch to the small black hole (SBH), and the other branch means there is an intermediate black hole. The intersection point of these branches marks critical point of the phase transition. Below the critical temperature, the LBH is more stable with lower free energy, and vice versa. The phase transition occurs as the temperature crosses this critical value. 
\par As indicated by the blue and orange line in \mpref{Q=0.032}, there is a black hole phase transition. However, for $Q>Q_c$ case, no phase transition is observed, as shown by the red and green lines. In the case of $Q<Q_c$, the size of black hole gradually increases with temperature until a critical point is reached, where two new black holes coexist with the existing black hole. As the temperature increases, one of these new black holes diminishes in size, while the other expands. These two black holes persist up to a certain temperature, beyond which the smaller and intermediate black holes coalesce and disappear, leaving only a large black hole at elevated temperatures. The occurrence of a phase transition precipitates a sudden shift in entanglement entropy, thereby changing both the form of Page curve and the Page time.
\par Then, We calculate the value of critical charge and the critical event horizon radius. To find these parameter, we take the partial derivative of the temperature equation
\begin{equation}
    \frac{d T}{d r_+}=\frac{1}{4 \pi}\left (-\frac{1}{ r_+^2}+3+\frac{3Q^2}{ r_+^4} \right),
\end{equation}
and make the temperature equation to be second-order derivative
\begin{equation}
    \frac{d^2 T }{dr_+^2}=\frac{1}{2\pi r_+^3}- \frac{3Q^2}{\pi r_+^5}.
\end{equation}
Setting the equation to zero allows us to establish the relationship $r_{+c}=\sqrt{6}Q_{c}$, from which we can deduce the critical electric charge $Q_c=0.167$. Consequently, we calculate the critical temperature $T_c$, which corresponds to the abscissa at the intersection point of the curves in \mpref{Q=0.031}. This critical point demarcates the boundary between different thermodynamic behaviors of the black hole.
\par
\begin{figure}[htb]
  \centering
  \includegraphics[scale=0.5]{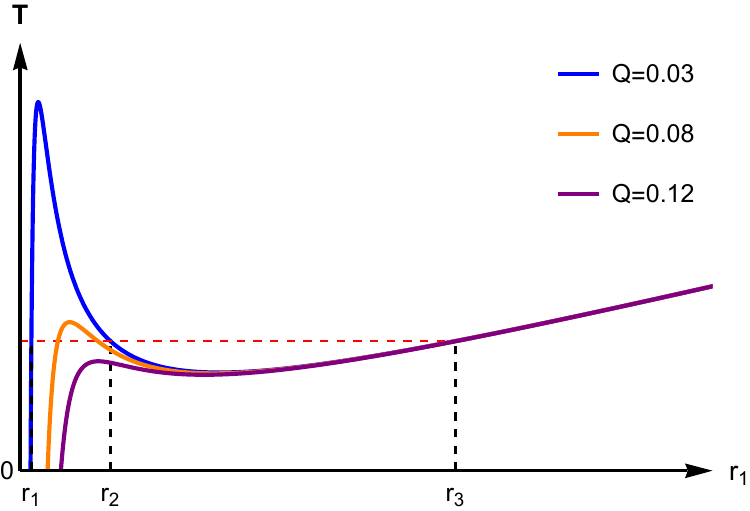}
  \caption{The $T-r_+$ diagram with $Q=0.03$, $Q=0.08$ and $Q=0.12$. There is only one black hole in the low temperature. As the temperature rises, the instability of the intermediate black hole leads to the existence of two distinct black holes.}
  \label{fig:r-T diagram}
\end{figure}

 In the aspect of thermodynamics, the black hole phase transition can be compared with the van der Waals first order phase transition mentioned before in \cite{kubizvnak2012p}. We plot the \mpref{Q=0.032} again, which added an ``isobar" and delineating three different value of $Q$ curves as \mpref{fig:r-T diagram} shows.
In the real evaporation process, for instance, the ``oscillating" part will be replaced by the part of red dotted line which corresponds to the temperature $T_0$ determined using the Maxwell equal area rule, in \mpref{fig:r-T diagram} when $Q=0.03$, and the critical charge is $Q_c=0.167$. As previously mentioned, the radius of black hole from $r_1$ to $r_3$ is unstable as mentioned before. The entanglement entropy will be elucidated as follows
\begin{equation}
        S = \text{Min}{ \left [ \frac{2}{3} \pi T(r_{+}) t,\ 2 S_{BH}  \right ] },
    \label{eq:Spiece}
\end{equation}
where $T(r_+)$ is defined by 
\begin{equation}
    T(r_+) = \begin{cases}
          \frac{1}{6 r_{+}} \left(1 + 3 r_{+}^2 - \frac{Q^2}{r_{+}^2} \right) , & \text{if } 0 < r_{+} < r_1 \\
          T_0 , & \text{if } r_1 < r_{+} < r_3 \\
          \frac{1}{6 r_{+}} \left(1 + 3 r_{+}^2 - \frac{Q^2}{r_{+}^2} \right) , & \text{if } r_{+} > r_3
        \end{cases}.
\end{equation}
\par Besides, the Page time varies with temperature. As the radius $r_+$ increases linearly, the Page time can be approximated by the following relationship 
\begin{equation}
        t_{\text{Page}}  \simeq \frac{3S_{BH}}{c \pi T(r_+)}.
\end{equation}
It is intriguing to examine the correlation between phase transition and Page time. The occurrence of a phase transition either before or after the Page time may influence the Page curve. This will be the subject of further discussion.
\par
\begin{figure}[htb]
\centering
\subfigure[\scriptsize{}]{\label{pt1}
\includegraphics[scale=0.35]{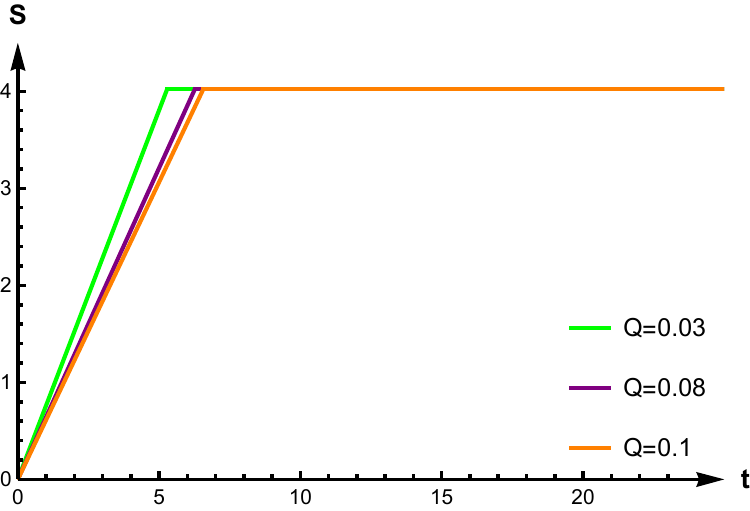}
}
\quad
\subfigure[\scriptsize{}]{\label{pt2}
\includegraphics[scale=0.35]{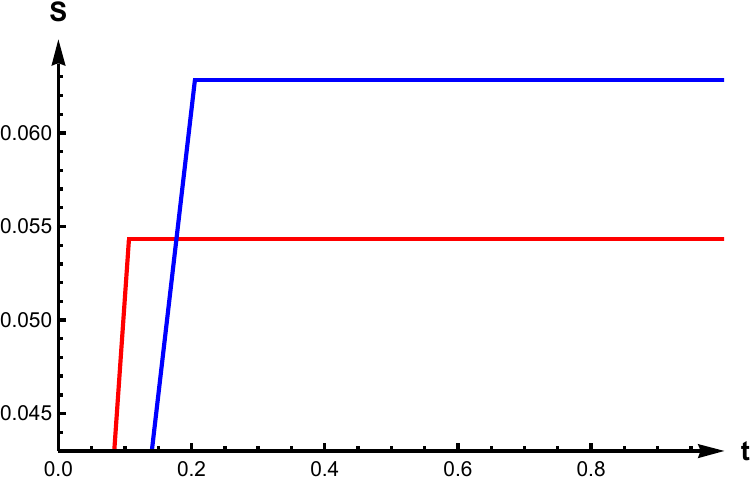}
}
\quad
\subfigure[\scriptsize{}]{\label{pt3}
\includegraphics[scale=0.35]{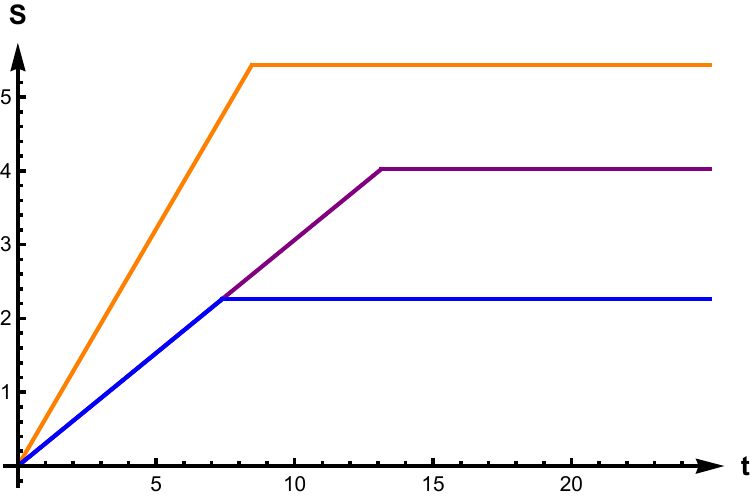}
}
\caption{The Page curves of eternal RN-AdS black hole. (a) Page curves for the charge $Q=0.03$, $Q=0.08$ and $Q=0.1$ cases, respectively. (b) Page curves for the charge $Q=0.08$. SBH with the red line, while intermediate black hole case with the blue line. (c) Page curves for the charge $Q=0.08$ case. Intermediate black hole case with purple and blue lines, and the LBH case with an orange line.}
\label{fig:Pc2}
\end{figure}

 In our study, we focus exclusively on the case where $Q<Q_c$ to investigate the impact of phase transitions. In the context of non-extremal eternal black holes, we consider three specific charge values: $Q=0.03$, $Q=0.08$, and $Q=0.1$. By substituting these values into the equation \eqref{eq:Spiece}, we can construct 3D Page curves. These curves are characterized by coordinates representing the event horizon radius of black hole $r_+$, time $t$, and generalized entropy $S$. 
\begin{figure}[htb]
\centering
\subfigure[\scriptsize{}]{\label{Q=0.033}
\includegraphics[scale=0.28]{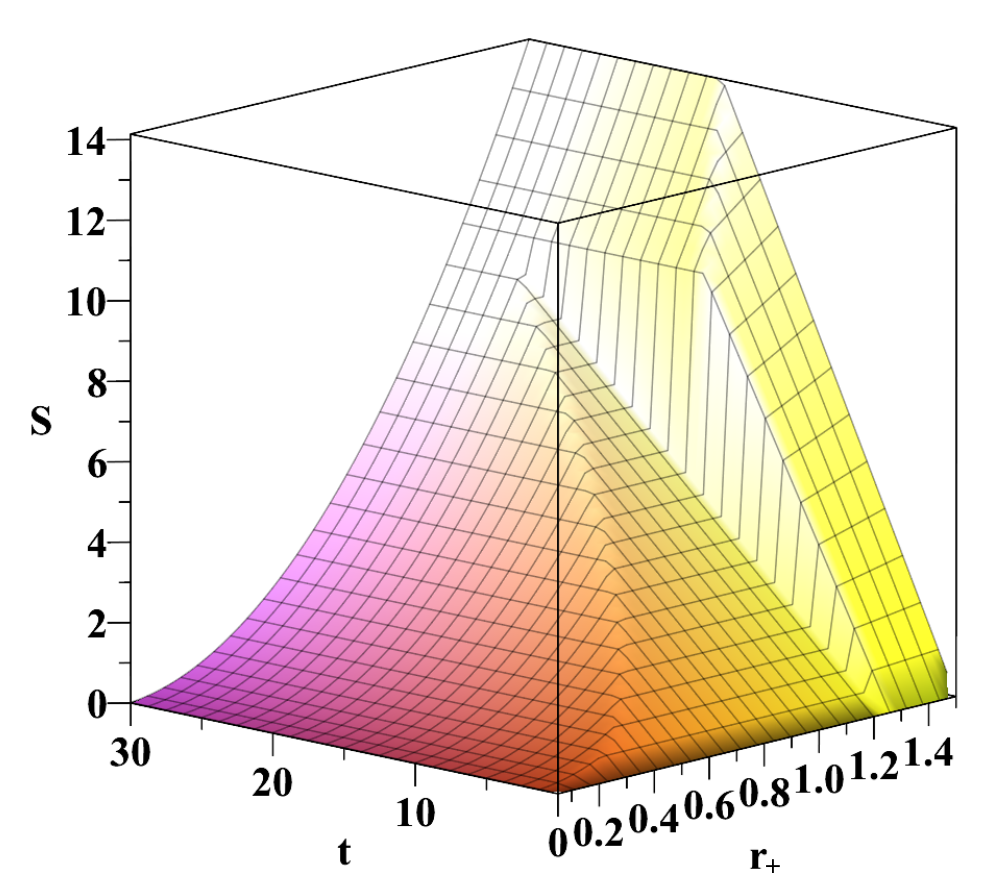}
}
\quad
\subfigure[\scriptsize{}]{\label{Q=0.034}
\includegraphics[scale=0.28]{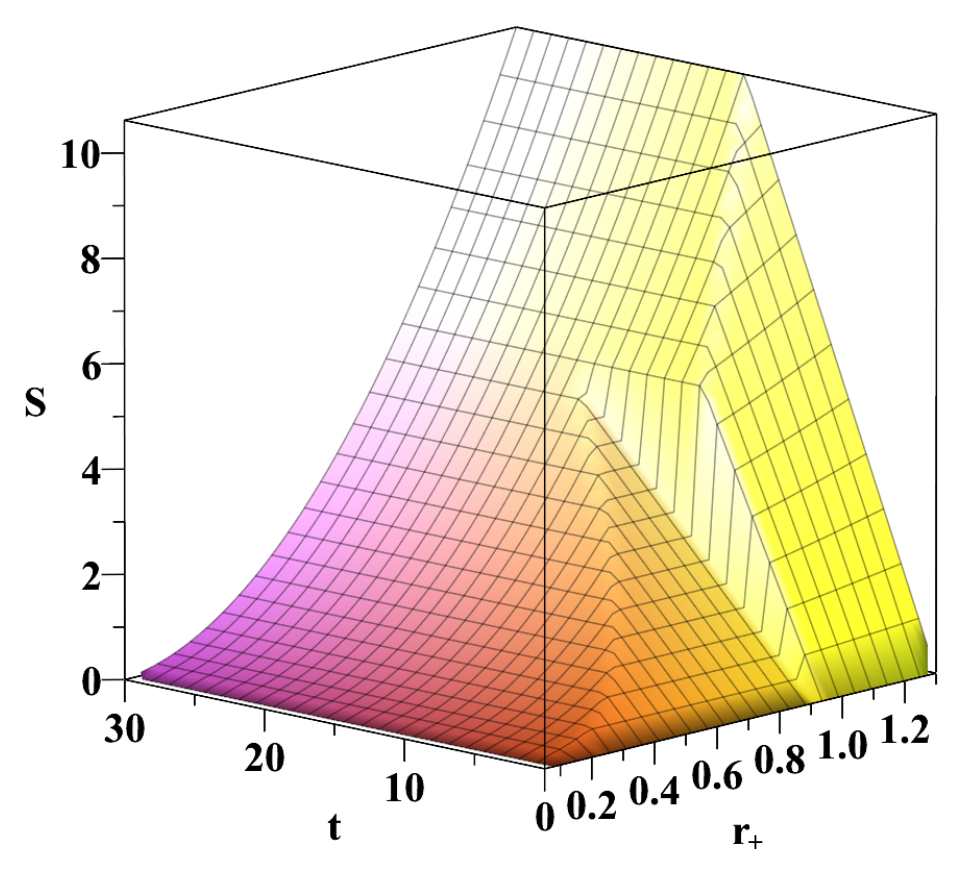}
}
\quad
\subfigure[\scriptsize{}]{\label{Q=0.035}
\includegraphics[scale=0.28]{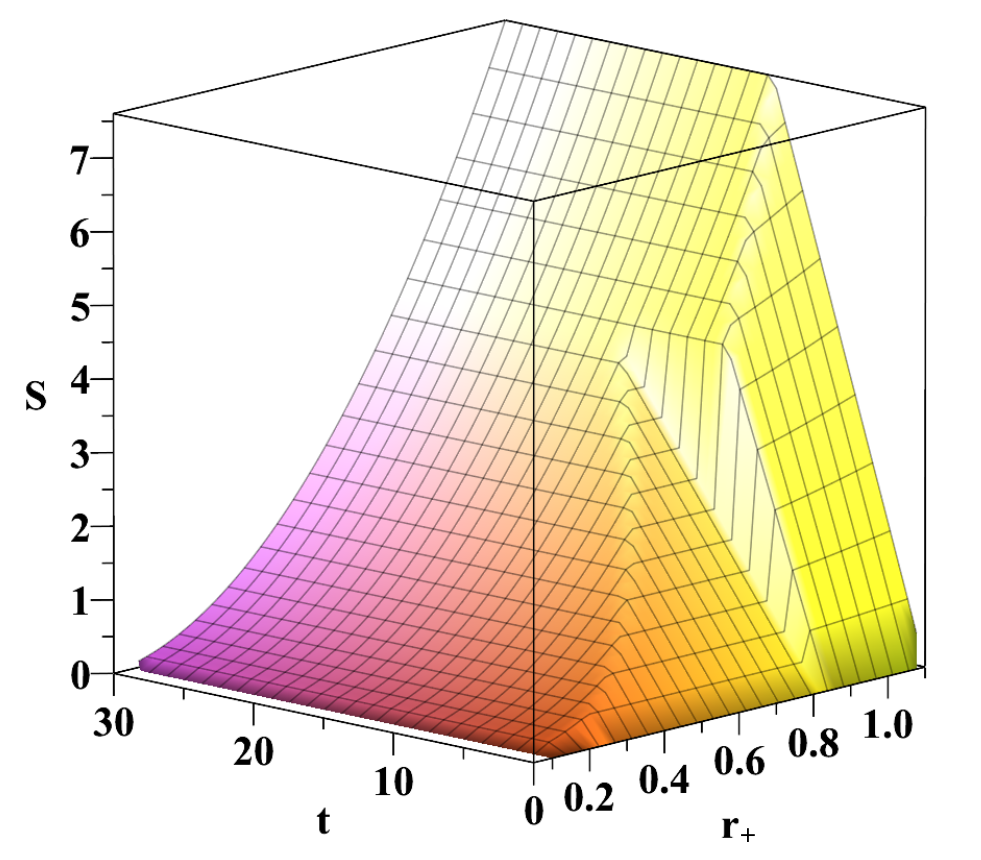}
}
\caption{The Page curves of eternal RN-AdS black hole. (a) Page curves for the charged $Q=0.03$ case. (b) Page curves for the charged $Q=0.08$ case. (c) Page curves for the charged $Q=0.1$ case.}
\label{fig:eternal}
\end{figure}
Analyzing these diagrams, one can observe that as the charge $Q$ decreases, the phenomenon of phase transition becomes more pronounced, which also as \mpref{pt1} shows. There are jumps between $r_{+}>r_{+c}$ and $r_{+}<r_{+c}$ which promise the phase transition to happen. When the event horizon radius $r_{+}$ is smaller than the critical value $r_{+c}$, the Page time exhibits a dependence on the charge $Q$. Specifically, it appears as a concave curve in the middle portions of the figures shown in \mpref{fig:eternal}. In the case of small $r_{+}$ (as depicted in \mpref{fig:r-T diagram} for $r_{+}<r_1$), the Page time occurs relatively early as figure \mpref{pt1} blue line shows.
However, as $r_{+}$ increases (corresponding to intermediate-sized black holes), the Page time shifts with $r_+$. Interestingly, for LBH, the Page time remains smaller than that of intermediate black holes but larger SBH case as shown in \mpref{pt2}. Conversely, when $r_{+}$ exceeds the critical value $r_{+c}$, the generalized entropy corresponding to the Page curve increases with $r_+$ (as described by \eqref{eq:Spiece}). Furthermore, at a fixed radius, smaller charge values lead to shorter Page times, as shown in \mpref{pt3}. Clearly, the charge $Q$ significantly influences both the Page time and the occurrence of phase transitions. Remarkably, we find that as the charge $Q$ decreases further, the impact on phase transitions becomes even more pronounced. 
\par Moving forward, we draw inspiration from previous works such as \cite{almheiri2019entropy,rocha2008evaporation}. Specifically, we explore the concept of coupling between one boundary and an external auxiliary system (acting as a heat sink). This coupling allows a two-sided black hole to undergo evaporation on one side. In the case of coupled baths, the black hole is allowed to exchange energy with its surroundings, leading to a gradual decrease in mass. The exchange process is assumed not to be influenced by the charge of black hole. As the black hole evaporates, its temperature and entropy undergo significant changes, which are reflected in the Page curve's behavior. Conversely, the removal of thermal baths isolates the black hole, causing it to follow a different evaporation trajectory. Without the energy exchange, the black hole's temperature rises sharply, accelerating the evaporation process. This rapid change can lead to a more pronounced effect on the Page curve, potentially resulting in a steeper decline or even discontinuities. Now we take it into consideration.
\par During the black hole’s evaporation process, its event horizon radius $r_+$ evolves with time. According to references \cite{page2005hawking,konoplya2019quasinormal}, the rate of change of $r_+$ with respect to time $dr_+/dt=-x$ is governed by a constant denoted as $x$. This constant incorporates fundamental physical quantities such as $\hbar$, $c$, $G$ and $M$.
\par
\begin{figure}[htb]
\centering
\subfigure[\scriptsize{}]{\label{BH1}
\includegraphics[scale=0.21]{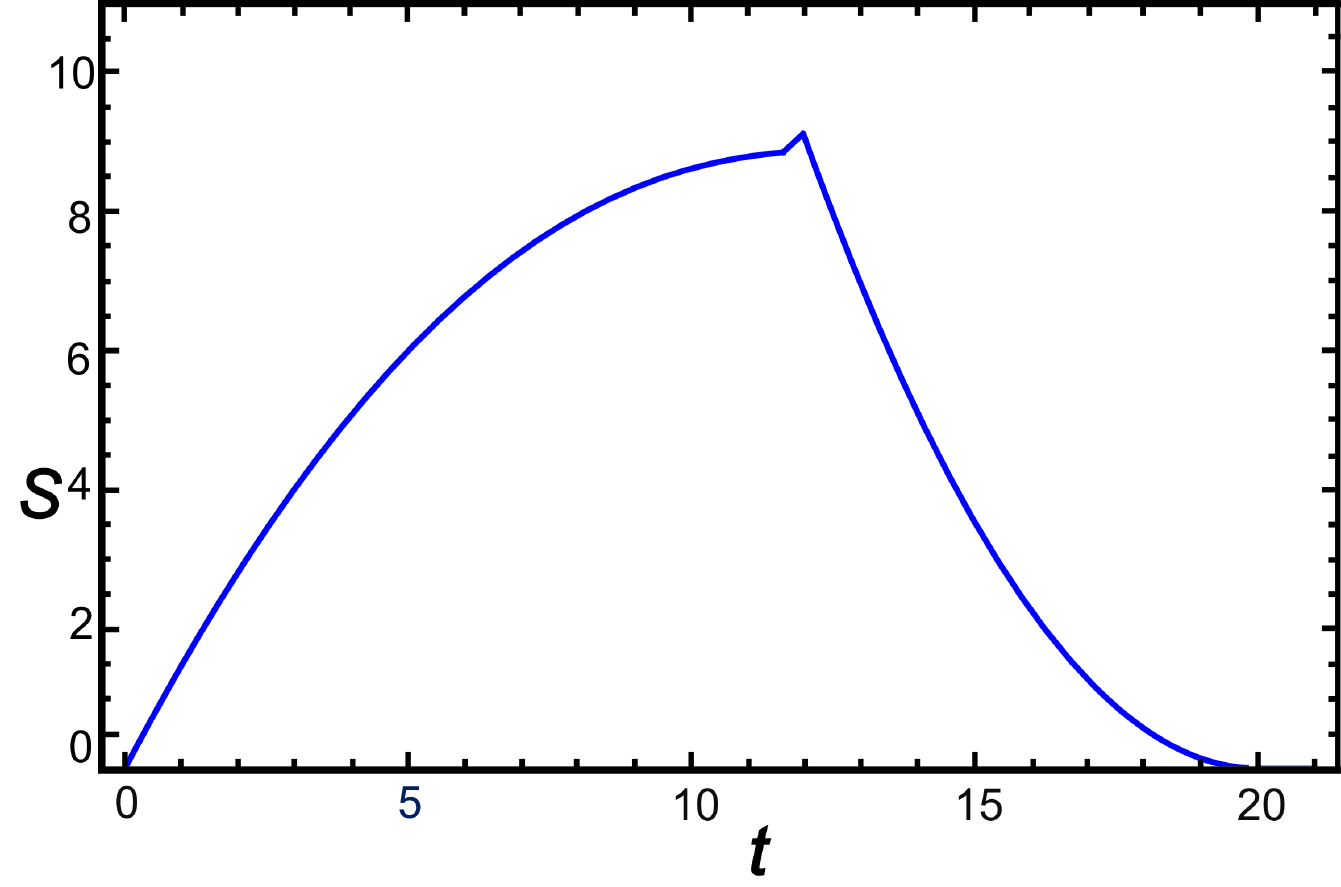}
}
\quad
\subfigure[\scriptsize{}]{\label{BH2}
\includegraphics[scale=0.21]{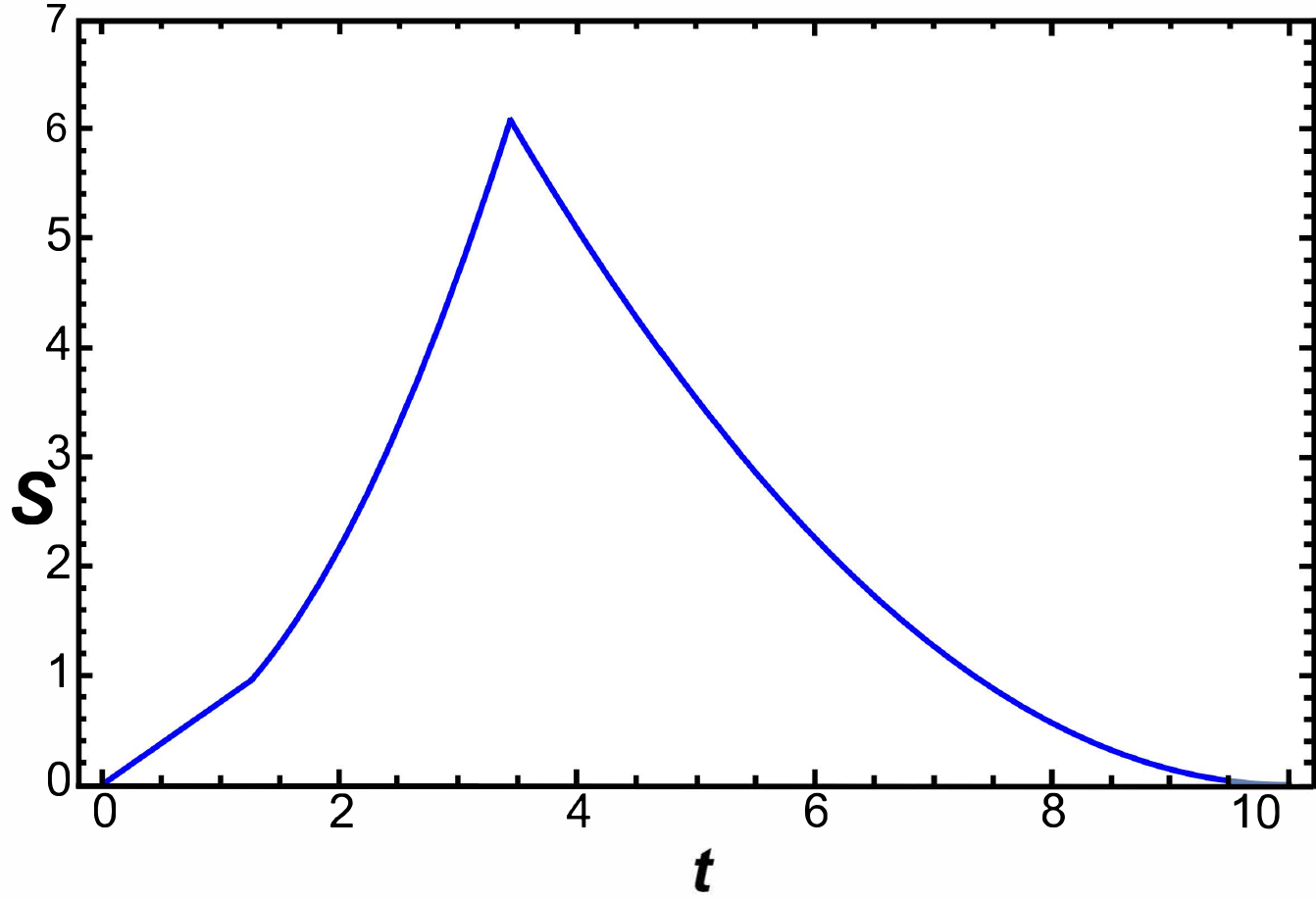}
}
\subfigure[\scriptsize{}]{\label{BH3}
\includegraphics[scale=0.21]{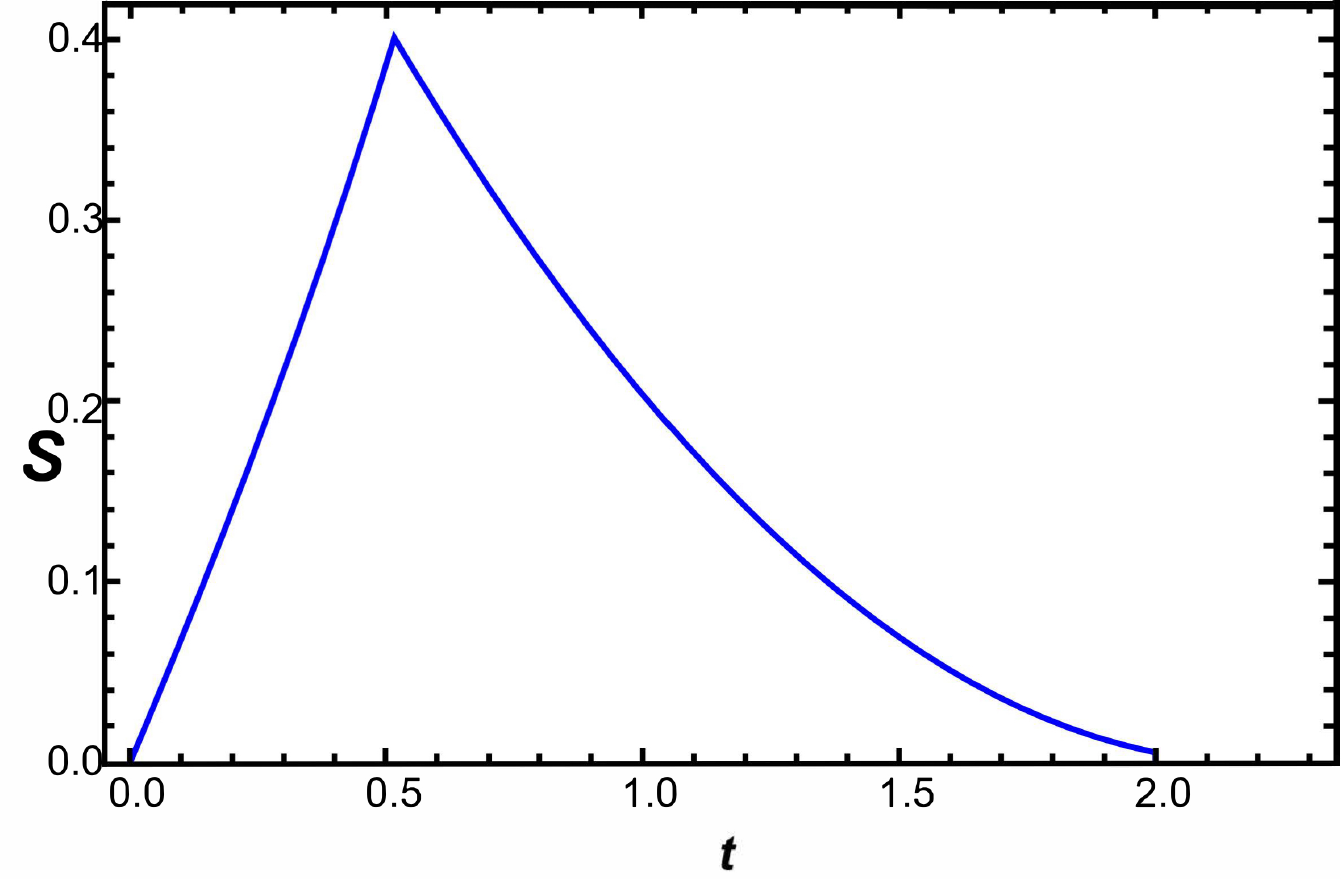}
}
\caption{The Page curve of a 4D RN-AdS black hole initially coupled to a bath, then isolated to allow for evaporation. (a) The Page curve for an RN-AdS black hole with a large horizon radius, where $r_{+}>r_1$, as shown in \mpref{fig:r-T diagram}. (b) The Page curve for an RN-AdS black hole with an intermediate event horizon radius. (c) The Page curve for a small RN-AdS black hole.}
\label{Page curve2}
\end{figure}

 For simplicity and clarity, we set the charge $Q=0.03$, ensuring that it remains smaller than the critical charge $Q_c$. This choice guarantees the occurrence of a phase transition during the process. The subsequent work discusses the Page curve structure for different ranges of the event horizon radius: $r>r_3$, $r_1<r<r_3$ and $r<r_1$. The Page curve diagrams corresponding to these three distinct event horizon scenarios are presented in \mpref{Page curve2}.
\par \mpref{BH1} shows that the Page curve initially increases in the early times. However, in the time half past the evaporation time, the Page curve for the LBH, where $r_{+}>r_1$, exhibits a discontinuity. Subsequently, the curve trends downward until the entropy reaches zero. If we assume the event horizon radius decreases linearly over time, the entanglement entropy reaches a minimum, and the Bekenstein-Hawking entropy term is depicted as a three-segment piecewise function. This configuration results in a Page curve that encapsulates the effects of the Hawking-Page phase transition, with the discontinuity arising from the phase transition itself. 
\par \mpref{BH2} demonstrates that for the intermediate black hole, which lies between $r_1$ and $r_2$ in \mpref{fig:r-T diagram}, the Page curve initially increases linearly. However, at the beginning of the evaporation, the discontinuity becomes more pronounced than in the Page curve of the large black hole. As with the larger black hole case, the form of the evaporating black hole is time-dependent. The Bekenstein-Hawking entropy term is modeled as a two-segment piecewise function, leading to an overall entropy that starts by following the minimal value of temperature times time $t$ and then shifts to a constant value, signaling a phase transition. This suggests that the unstable black hole transitions into a smaller black hole before the phase transition. The sharp change in temperature as the black hole's phase shifts causes the slope changes, indicating that if the phase transition occurs before the Page time, the Page curve will exhibit a notable deviation.
\par \mpref{BH3} depicts the Page curve of SBH is monotonically increasing, but it declines after the Page time. At late stage time of the process, the entanglement entropy turns into zero suddenly. In the final stages of the evaporation process, the entanglement entropy abruptly reduces to zero. As depicted in \mpref{fig:r-T diagram}, the absence of a phase transition maintains the Page curve in its canonical form. Consequently, the Page curve for a small black hole does not reflect any phase transition. Notably, for minuscule radii, the Page time exhibits an increase over time in contrast to that of a larger black hole, attributable to the negative heat capacity.

\section{Discussion and Conclusion} \label{Discussion}
\qquad We study the Page curve of Reissner-Nordström AdS black holes in 4D spacetime under a variety of conditions, by using the island formula \eqref{Island}. Initially, we introduce couplings to the bath, enabling Hawking radiation to reach null infinity and initiating the evaporation of the RN-AdS black hole. In scenarios without island, entropy escalates indefinitely over time, as delineated in Equation \eqref{eq:without island}, contradicting the Bekenstein-Hawking entropy bound and violating the unitarity principle of quantum mechanics. To resolve this paradox, we integrate the concept of an entanglement island in the RN-AdS black hole framework. Our subsequent analysis of the process involving an island reveals that multiple islands yield a smoother Page curve compared to the single island case. According to the island formula, after the Page time, the Bekenstein-Hawking entropy is the predominant term. After the Page time, the entanglement entropy stabilizes at a constant value due to the emergence of islands, as calculated in Equation \eqref{eq:with island}, reflecting the behavior of an evaporating black hole. The Page curve depicting the evaporation process of the eternal black hole is illustrated in \mpref{fig: Page curve}.
\par Next, we explore the impact of the charged parameter $Q$ on the Page curve when $Q<Q_c$. Specifically, we plot the Page curve for the eternal 4D RN-AdS black hole, as shown in Figure \ref{fig:eternal}. Remarkably, we observe that as the charge $Q$ decreases, the effects on the Page curve become more pronounced, as shown in \mpref{pt2}. During phase transitions, the early stages of Hawking radiation entropy exhibit significant changes. In the context of replica wormholes, the transition from the Hawking saddle to the replica wormholes saddle involves additional time, which indicated in the \mpref{fig:eternal}, where the curve spends more time reaching to Page time. However, when no phase transition occurs, the Page curve aligns with the original proposal by Page, where the generalized entropy depends on the value of the event horizon radius $r_+$. We note that the charge factor becomes pivotal when a charged black hole evaporates.
\par We assume that the radius monotonically decreases over time, implying that Hawking radiation does not deplete the black hole's charge, thus maintaining a constant charge throughout the process. Through an examination of the island configuration and phase transitions, we investigate information preservation in 4D RN-AdS black holes, focusing on the effects of phase transition on the Page curve across various black hole sizes. The large black hole case shows an initial increase in entanglement entropy, following the Page curve, but displays a discontinuity at the Page time due to a phase transition, as indicated in \mpref{BH1}. The intermediate black hole scenario exhibits minimal phase transition effects, with unstable black holes transitioning to smaller phases before the Page time, as marked in \mpref{BH2}. The SBH scenario experiences no phase transition, and its Page curve follows the typical pattern of an evaporating black hole, as shown in \mpref{BH3}. This anomaly phenomenon leads us to consider the preservation of the information charge $Q$ in the black hole evaporation process. In the mainstream research of the black hole information paradox, information is not lost. Thus, the charge should undergo some physical process to determine its residue or evaporation.
\par We find that the discontinuity observed in \mpref{BH1} could indicate a physical event, potentially neutralizing the charge and transforming the black hole into an evaporating Schwarzschild black hole, which ultimately dissipates entirely. This process could also influence the replica trick used to compute entropy, as it may prolong the duration on the Hawking saddle prior to transitioning to the wormholes saddle. Additionally, one can consider the effect of charge $Q$ in the path integral of the replica trick. Such an information problem presents an intriguing avenue for future research.
\par We conclude that phase transitions indeed impact the Page curve. These transitions may occur before or after the Page time, depending on the initial size of the black hole. Nonetheless, these results do not compromise the unitarity principle of the quantum system, indicating that the paradox can be resolved under these conditions. A prospective research direction could involve examining the mechanisms by which the charge dissipates to zero in RN-AdS black holes coupled with a bath. One interesting aspect is that entanglement entropy might have a relation with the heat engine. The Hawking saddle and replica wormholes saddle may correspond to the liquid or gas phase transition. The parameter to explain the real phase transition could be Renyi entropy or others.

\section*{Acknowledgement}
\qquad We would like to thank YuQi Lei, ChengYuan Lu and ChenYang Dong for helpful discussions. This work is partly supported by the National Natural Science Foundation of China (Grant No.12275166 and No. 12311540141).

\end{document}